\documentclass[preprint,12pt]{elsarticle}

\usepackage{amssymb,gensymb,url,amsmath}

\journal{Physica A}

\begin{document}

\begin{frontmatter}



\title{Eliminating Bias in Pedestrian Density Estimation: A
Voronoi Cell Perspective}


\author[inst1,inst2]{Pratik Mullick}

\affiliation[inst1]{organization={Department of Operations Research and Business Intelligence, Wrocław University of
Science and Technology},
            addressline={Wyb. Wyspiańskiego 27}, 
            city={Wrocław},
            postcode={50370}, 
            state={Lower Silesia},
            country={Poland}}

\affiliation[inst2]{organization={Univ Rennes, INRIA, CNRS, IRISA},
            addressline={263 Av. Général Leclerc}, 
            city={Rennes},
            postcode={35042}, 
            state={Bretagne},
            country={France}}

\author[inst3]{Cécile Appert-Rolland}
\affiliation[inst3]{organization={Université Paris-Saclay, CNRS/IN2P3, IJCLab},
            addressline={15 Rue Georges Clemenceau}, 
            city={Orsay},
            postcode={91400}, 
            state={Île-de-France},
            country={France}}
            
\author[inst4]{William H. Warren}
\affiliation[inst4]{organization={Department of Cognitive, Linguistic and Psychological Sciences, Brown University},
            addressline={190 Thayer Street}, 
            city={Providence},
            postcode={02912}, 
            state={Rhode Island},
            country={USA}}

\author[inst2]{Julien Pettré}

\begin{abstract}
For pedestrians moving without spatial constraints, extensive research has been devoted to develop methods of density estimation. In this paper we present a new approach based on Voronoi cells, offering a means to estimate density for individuals in small, unbounded pedestrian groups. A thorough evaluation of existing methods, encompassing both Lagrangian and Eulerian approaches employed in similar contexts, reveals notable limitations. Specifically, these methods turn out to be ill-defined for realistic density estimation along a pedestrian’s trajectory, exhibiting systematic biases and fluctuations that depend on the choice of parameters. There is thus a need for a parameter-independent method to eliminate this bias. We propose a modification of the widely used Voronoi-cell based density estimate to accommodate pedestrian groups, irrespective of their size. The advantages of this modified Voronoi method are that it is an instantaneous method that requires only knowledge of the pedestrians’ positions at a give time, does not depend on the choice of parameter values, gives us a realistic estimate of density in an individual’s neighborhood, and has appropriate physical meaning for both small and large human crowds in a wide variety of situations. We conclude with general remarks about the meaning of density measurements for small groups of pedestrians.
\end{abstract}



\begin{keyword}
pedestrian dynamics \sep crowd management \sep density estimation \sep voronoi construction 
\end{keyword}

\end{frontmatter}


\section{Introduction}\label{sec:intro}

Human crowd motion studies \citep{sweeny2013perceiving,haghani2018crowd,warren2018collective,WANG2019domino,HAGHANI2021,SYED2022,WANG2023fire,CHEN2024} focus on understanding and modeling \citep{duives2013state,van2021algorithms} the behavior of individuals within large groups as they move through and interact with their environment. These studies are crucial for designing safer public spaces, optimizing evacuation procedures, and improving crowd management during events \citep{zheng2009modeling,ma2016effective,bakar2017overview,li2019review,CHEN2023129002}. Human crowd motion is an example of a complex system because it involves numerous interacting agents (people) whose collective behavior cannot be easily predicted from the actions of individuals alone. Factors such as collective decision-making \citep{dyer2008consensus,SAAKIAN20181408,LIU2022127087}, social influences \citep{lorenz2011social,becker2017network,frey2021social,MAVRODIEV2021125624}, spatial constraints \citep{pournajaf2014spatial,abousamra2021localization}, and environmental conditions \citep{haghani2018crowd,filingeri2017factors} lead to emergent phenomena, such as congestion \citep{helbing2005,king2014using,HUANG2015200,Feli2016pre,feliciani2018measurement,zanlungo2023pure,KAWAGUCHI2023128547}, lane formation \citep{GUO201587,cao2017,ZHANG201972,BODROVA2024129796,FANG2024129626}, and phase transitions between different flow states \citep{muramatsu2000jamming,nagatani2002dynamical,tajima2002clogging,nagatani2009freezing,LIU2024129499}. This complexity arises from the nonlinear interactions and adaptive behaviors exhibited by individuals, making crowd motion a rich field for exploring the principles of complex systems.

In the context of self-organizing behaviour of human crowd motion, methods of density estimation are an important topic of research \citep{seyfried2005,helbing2007pre,seyfried2010,schauer2014estimating,DUIVES2015162,rao2015estimation,saleh2015recent,tordeux2015,Nagao2018,ding2020crowd}. The existing literature consists of a number of methods, although the `best’ method is unresolved and may depend on the crowd situation. Most of the research has focused on situations in which the moving crowd is constrained within a physical boundary, such as a corridor or sidewalk \citep{sankaran2016,dali2016}. For such cases, density estimation using a Voronoi tesselation \citep{seyfried2010,zhang2011,DUIVES2015162,cao2017,cecile_vor}and a grid-based measure called the XT method \citep{EDIE1963,bode_c_h2019,saberi2014} have been reported to work well. However, there is no well defined method of density estimation for groups in an unbounded space.

For efficient crowd management, a crucial aspect is the construction of a fundamental diagram (FD) \citep{WANG2019,Lian2022,PAETZKE2022,Cristiani2023,ZENG2023,RANGELGALVAN2024}. In the context of traffic flow, an FD basically depicts the relationship between traffic velocity and traffic density. This relationship could help to study the capacity of a space where the traffic moves, e.g., a road for vehicles, sidewalks for pedestrians. An FD could serve as a basic element of comprehensive models that describe the traffic operation on a network, thereby finding significant applications in the context of human traffic management and crowd safety \citep{GEROLIMINIS2008759,KEYVANEKBATANI20121393}. The FD could also be used as a valuable tool for assessing the capacity of pedestrian flow simulations to accurately predict real-world scenarios. Therefore, the construction of a realistic FD necessitates the utilization of an effective method for density estimation.

A further difficulty arises when the fundamental diagram, rather than being plotted at a global or meso scale \citep{seyfried2005}, is related to individual quantities \citep{jelic2012a}. In one dimension, a proxy for density can be the inverse of the spatial headway with the predecessor. However, in two dimensions, finding a robust estimate of the density along the trajectory of a pedestrian is significantly more complex. Estimating local density in two-dimensional spaces requires capturing the spatial interactions among pedestrians accurately. The existing methods of density estimation often fall short in scenarios where pedestrian groups move in unbounded spaces or when the group size is relatively small. Our research introduces a density estimation method tailored to these specific conditions.

In \citep{cecile_vor} the authors developed a Voronoi cell based density estimation method for unbounded pedestrian groups in which the Voronoi cells are restricted to the angular sectors on the convex hull of the whole set of pedestrians. However, that method was demonstrated only for pedestrian groups larger than $\sim 40$ individuals. In our current paper we point out some typical cases that could arise when the pedestrian group is much smaller ($\sim 5$). Our main contribution is a technique by which we can modify the Voronoi method to adjust the angular corrections for such small groups, and obtain a realistic estimate of the density felt by an individual. With our added modification, the computational algorithm becomes more general and applicable to pedestrian groups of any size, even when the number of pedestrians is as small as 5.

In practical terms, this means that our method can dynamically adapt to the spatial distribution and movement patterns of pedestrians, offering a more realistic and bias-free estimate of local density. This is particularly important in unbounded environments, where traditional methods may struggle to account for the lack of spatial constraints. Our approach ensures that the density felt by an individual pedestrian is accurately captured, providing useful insights into crowd dynamics and interactions at a microscopic scale.

Furthermore, our method has the potential to enhance the construction of fundamental diagrams by providing precise density measurements at the individual level. This allows for a more detailed analysis of pedestrian behavior and flow characteristics, facilitating the development of more effective crowd management strategies and safety measures. By addressing the limitations of existing density estimation techniques, our proposed method represents a significant advancement in the study of human crowd motion and contributes to a deeper understanding of the complexities inherent in pedestrian dynamics.

The rest of the paper is organized as follows: in the next section we briefly describe the data set that has been used in this research to demonstrate our proposed computational algorithm. Then we briefly describe the existing methods of density estimation in the literature, and our proposal for adapting the Voronoi method for small groups without spatial boundaries. In the Results and Discussion section, we present an extensive evaluation of the earlier methods, which highlights their limitations and drawbacks. Then we show how our proposed method outperforms the previous methods in providing a bias-free realistic estimate of density in the neighborhood of an individual in small groups. We conclude with a general discussion on the meaning of defining a density at such small scales.

\section{Materials and Methods}\label{sec:mat.method}

\subsection{Experimental details}\label{sec:expt}

\begin{figure}[h!]
    \centering
    \includegraphics[width=\linewidth]{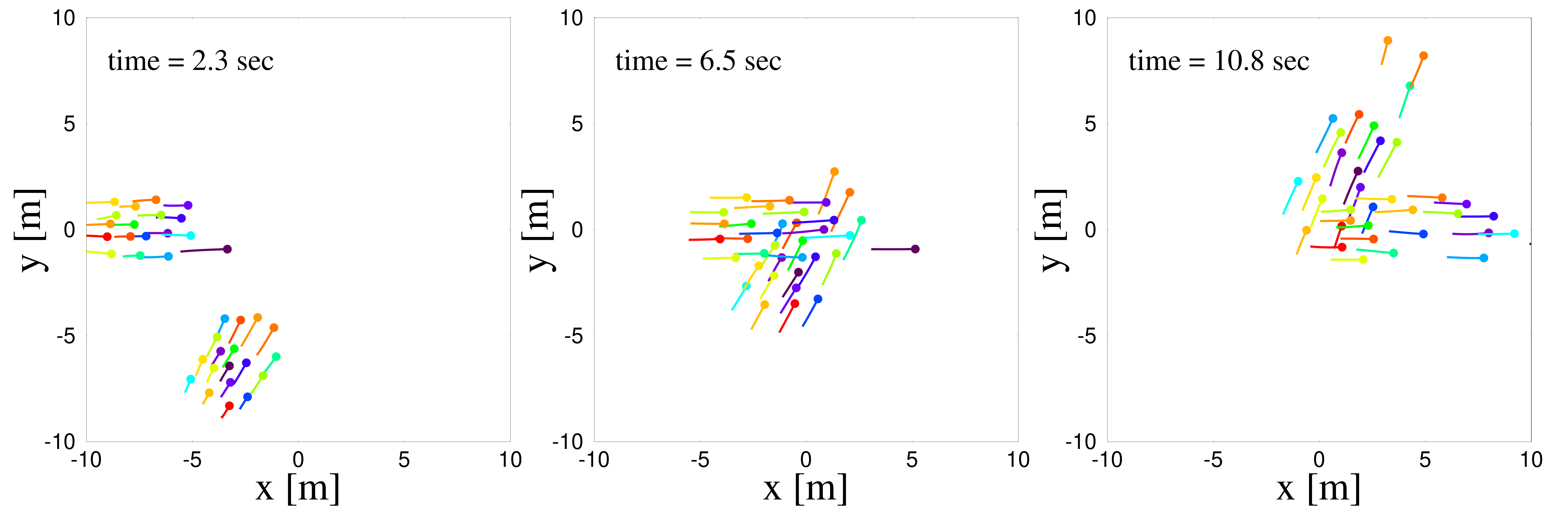}
    \caption{Displacements ($1.25$ sec) of all pedestrians in a typical trial with an expected crossing angle of $60\degree$ . The displacements are shown at three different time points, viz. $2.3$ sec, $6.5$ sec and $10.8$ sec, from the beginning of the trial. The dots represent the pedestrians and the tails behind each dot are the distances travelled by the pedestrian in previous $1.25$ sec.}
    \label{fig:traces}
\end{figure}

For this research, we consider the typical situation of crossing flows of pedestrians without any spatial constraints, where two groups of people walk across an open area from predefined initial positions such that their paths cross each other at specified values of the crossing angle. The data were obtained by experiments \citep{pedinteract_cecile,ploscb_pratik} using live participants on the campus of University of Rennes, France. This data is available in a public repository \url{https://doi.org/10.5281/zenodo.5718430}. Two different sets of volunteered participants (36 on Day 1, 38 on Day 2) were roughly divided into two groups (18 or 19 per group) and were instructed to reach the other side of a sport hall. Initial positions were prescribed so that the groups had to cross each other with seven different crossing angles ($0\degree$ to $180\degree$ , at $30\degree$ intervals). During each trial we recorded the head trajectory of each pedestrian as a time series at a frequency of 120 Hz using a motion capture system based on infrared cameras (VICON). The data obtained were then low-pass filtered to decrease the oscillations due to the gait movement of the walking pedestrians. Precisely, we used a forward-backward 4th-order butterworth filter to these unwanted oscillations with a cut-off frequency of $0.5$ Hz. In Figure \ref{fig:traces} we show the traces of all the pedestrians for a typical trial using filtered trajectories. 

\subsection{Methods of density estimation}\label{sec:methods}

A large number of methods have been developed to measure the density field of pedestrian flows \citep{seyfried2010,DUIVES2015162}. In order to determine the fundamental diagram at an individual scale, one has to associate a density with a specific pedestrian location. In this subsection we briefly present three such methods which already exist in the literature in the context of pedestrian flows, followed by the Voronoi method \citep{cecile_vor} that we have modified to be applicable for pedestrian groups irrespective of their size. In the next section we shall evaluate the effectiveness of these methods.

\subsubsection{Grid-based Classical Method}\label{sec:grid}

The classical method to estimate the density of pedestrians follows an Eulerian approach. In this method we divide the entire tracking region into a grid of square cells. If $d_g$ is the length of one side of a square cell, the classical density $\rho_g$ is given by \begin{equation}
    \rho_g=\frac{n}{d_g^2}\label{eq:grid}
\end{equation} where $n$ is the number of pedestrians that are located inside the square cell. The density estimated for each of the square regions is associated with all the pedestrians that are within the square. Note that sometimes, a single cell is used to measure density in a region of interest, such as in the crossing area of crossing flows \citep{guo2010,zanlungo2023a,zanlungo2023b}.

\begin{figure}[h!]
    \centering
    \includegraphics[width=\linewidth]{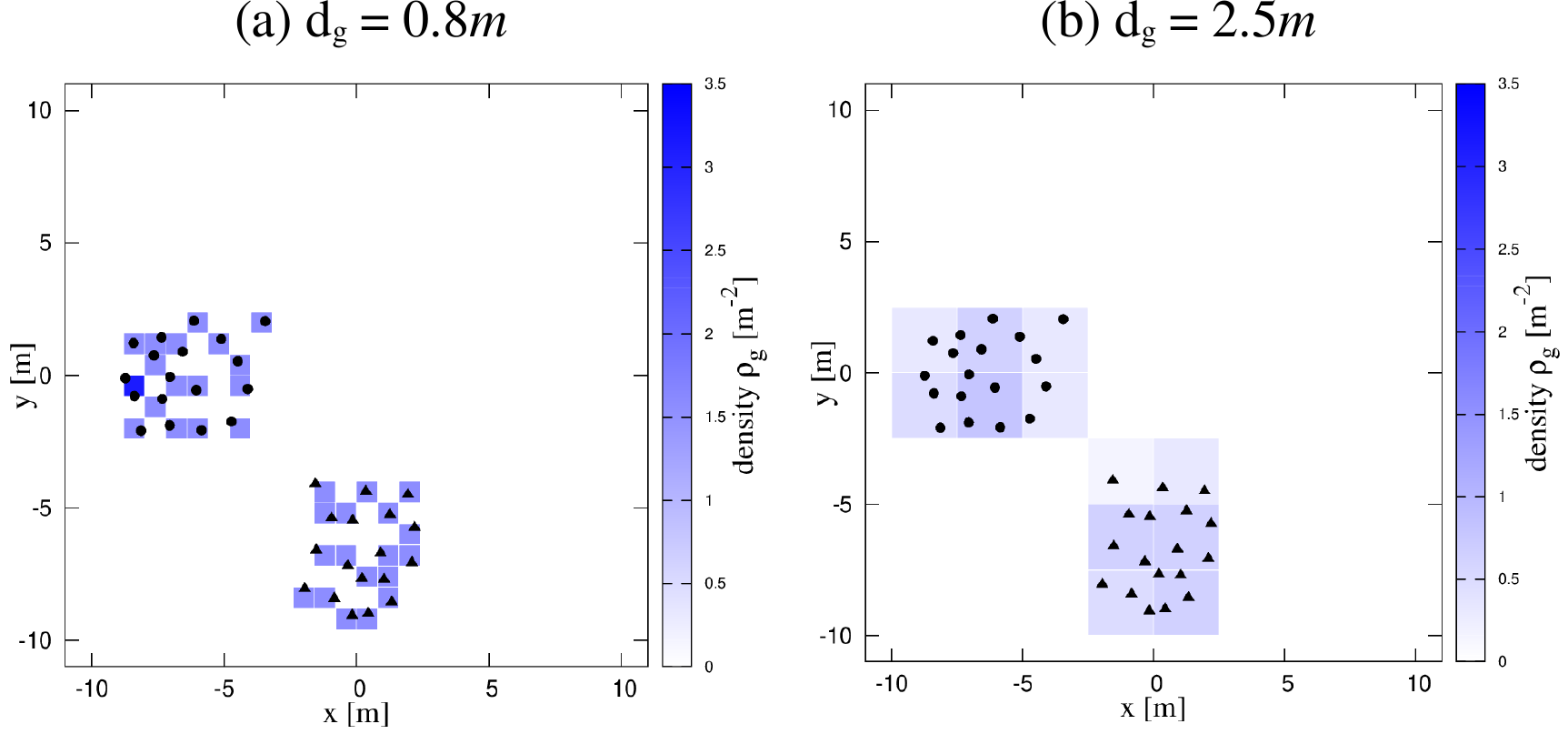}
    \caption{Density $\rho_g$ fields as estimated using the grid-based classical method for two typical values of $d_g$ , viz. (a) $d_g = 0.8$m and (b) $d_g = 2.5$s. Density values decrease with the increase in $d_g$ . For this demonstration we use a typical data from our crossing flows data set. The two groups of people, denoted by black circles and triangles, attempt to cross each other at $90\degree$. The group denoted by circles move along the $x$-axis from negative to the positive direction, whereas the group denoted by triangles move along the $y$-axis from negative to positive direction.}
    \label{fig:fields_grid}
\end{figure}

In Figure \ref{fig:fields_grid} the density $\rho_g$ fields of the pedestrians for a typical case in our data set has been shown for two typical values of $d_g$.  Clearly the density fields show variation with variation in $\rho_g$ . Precisely, as $d_g$ increases, density $\rho_g$ values decreases, as also evident from Equation (\ref{eq:grid}). In Figure 9(a) we show the time-sequence of classical density $\rho_g$ for a typical pedestrian for several values of the cell size. For smaller cell sizes, the density values keep switching between only a few levels - which signifies the discrete nature of $\rho_g$.

\subsubsection{XT Method}\label{sec:xt}

The XT method was originally proposed by Edie \citep{EDIE1963} in the context of traffic stream measurements, where the total distance travelled and the total time spent by the pedestrian in a space-time region is taken into account. Edie’s definition was extended \citep{saberi2014} to be applicable to a three-dimensional space for multi-directional pedestrian motion. Later, this method was further modified \citep{DUIVES2015162}, which we have used in this paper to compute the pedestrian density.

\begin{figure}[h!]
    \centering
    \includegraphics[width=0.65\linewidth]{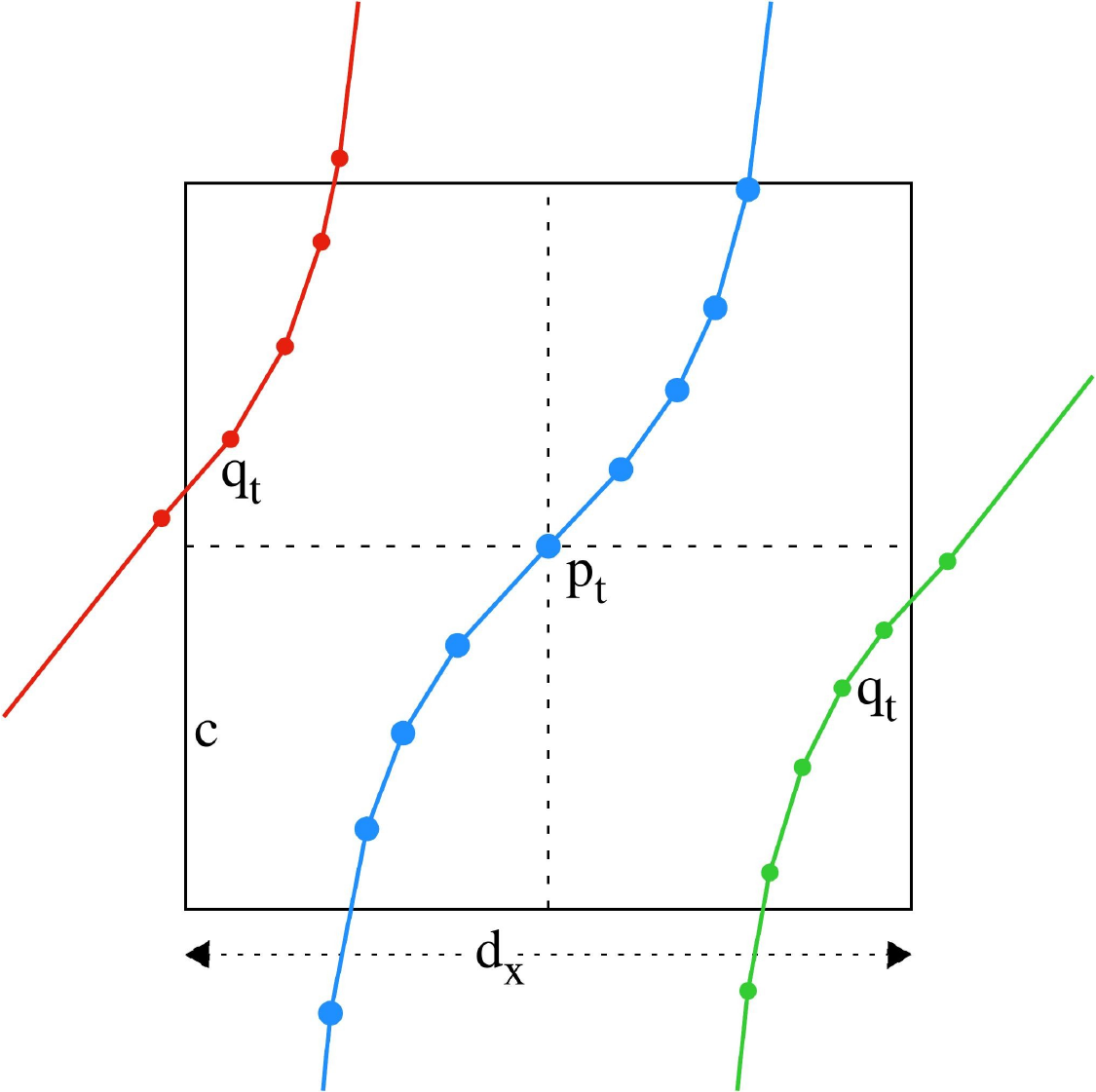}
    \caption{Schematic diagram showing the XT method to calculate the pedestrian density $\rho_\text{xt}$. The density calculated is associated to the pedestrian $p_t$ , whose trajectory is shown in blue. At time $t$ this pedestrian is located at the centre of the square shaped cell $c$ of length $d_x$. Other pedestrians $q_t$, whose trajectories are shown in red and green also reside within this cell at time $t$ and therefore are considered for density evaluations.}
    \label{fig:schematic_XT}
\end{figure}

\begin{figure}[h!]
    \centering
    \includegraphics[width=\linewidth]{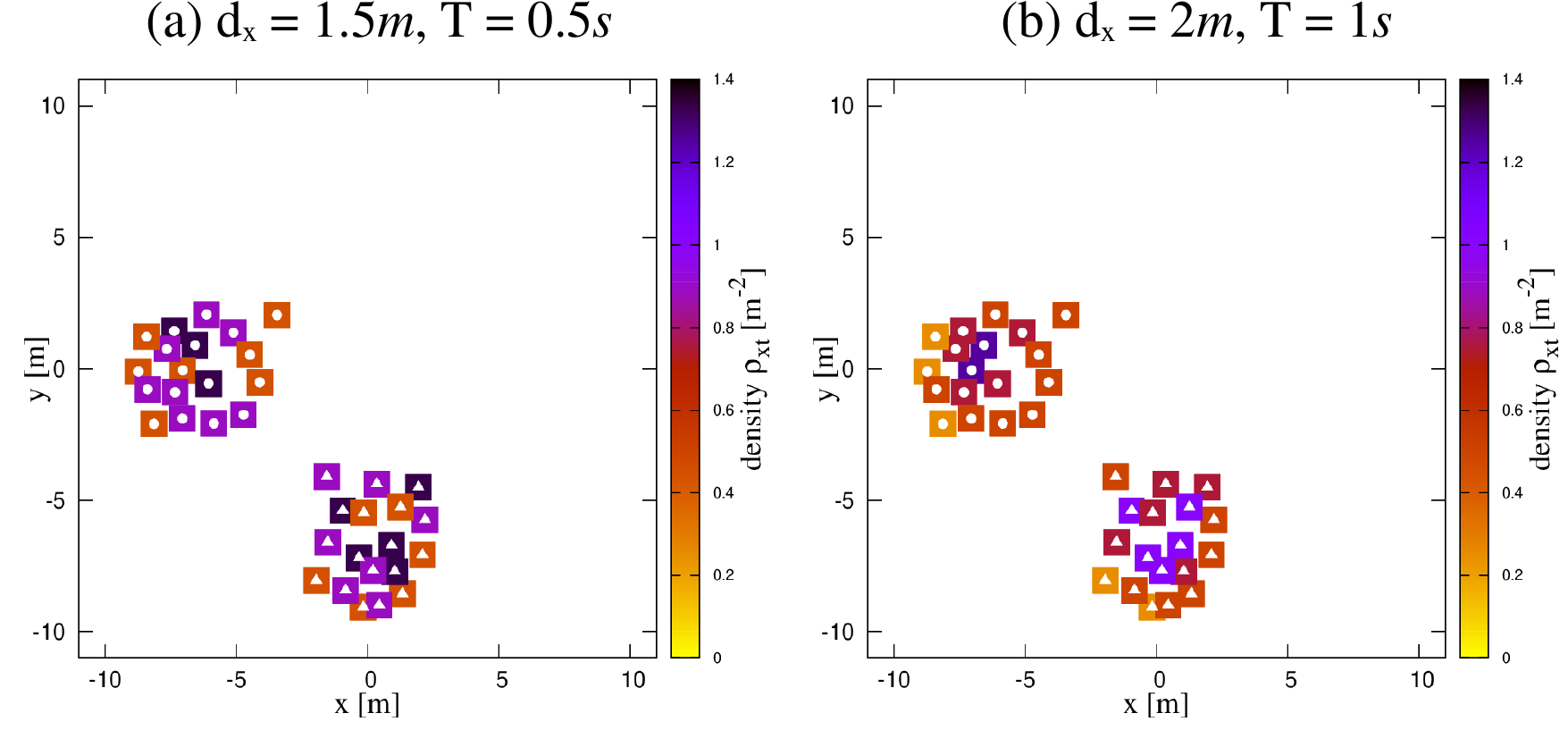}
    \caption{Density $\rho_\text{xt}$ fields as estimated using the XT method for two sets of typical values of $d_x$ and $T$ , viz. (a) $d_x = 1.5$m, $T = 0.5$s and (b) $h = 2$m, $T = 1$s. Density values decrease with the increase in $d_x$ and $T$. The data used here is the same one that has been used in Figure \ref{fig:fields_grid}.}
    \label{fig:fields_xt}
\end{figure}

A square-shaped cell $c$ of length $d_x$ is considered at whose center a pedestrian $p$ is located at time $t$, as shown in Figure \ref{fig:schematic_XT}.  With the progression of time, this cell travels with $p$, yielding the local density along the trajectory of $p$. This pedestrian-centered frame of reference contrasts with the classical method described previously, which has a space-centered frame of reference. There could be other pedestrians as well within this cell at time $t$. The density $\rho_\text{xt} (c, t)$ estimated for this cell at time $t$ is associated with the pedestrian $p$ and is given by \begin{equation}
    \rho_{\text{xt}}(c,t)=\frac{\sum_q(T^q_{\text{end}}-T^q_{\text{begin}})}{{d_x}^2 \times T}\label{eq_xt}
\end{equation} where the summation is performed over all the pedestrians $q$ placed inside the cell $c$ at time $t$. $T^q_\text{begin}$ denotes either the time when the pedestrian $q$ enters the cell $c$ or the lower time boundary $t - 0.5T$, whichever is minimum and $T^q_\text{end}$ denotes the minimum of the time when the pedestrian $q$ exits the cell $c$ and the upper time boundary $t + 0.5T$. $d_x^2$ is the area of the cell $c$. The timescale $T$ restricts the time window of the pedestrians within which they are accounted for during the computation. In Figure \ref{fig:fields_xt} the density $\rho_\text{XT}$ fields of the pedestrians for two sets of typical values of $d_x$ and $T$ have been shown, using a typical data from our crossing flows data set. 

\subsubsection{Kernel Method}\label{sec:kernel}

To estimate the density of pedestrians in a two-dimensional unbounded space we also use a non-parametric method based on the kernel density estimator. This method basically evaluates the probability density function of the pedestrian positions, from which we calculate the density of pedestrians. Following a Lagrangian approach, this method gives an estimate of density even at the spatial positions where there are no data points.

For $\textbf{X}_1,\textbf{X}_2,\textbf{X}_3,\dots,\textbf{X}_N$ to be the collection of two-dimensional coordinates
for $N$ pedestrians, the density function $\rho_k$ estimated by the kernel density method at the spatial position $\textbf{X}$ is given by
\begin{equation}
    \rho_\text{k}^{\textbf{h}}(\textbf{X}) = \sum^{N}_{i=1} K_{\textbf{h}}(\textbf{X}-\textbf{X}_i),\label{eq:kernel}
\end{equation} where $\textbf{h}$ is the bandwidth which dictates the smoothness of the density measurement. Among the several choices of the kernel function $K$, we use the bi-variate Gaussian distribution function given by, \begin{equation}
    K_{\textbf{h}} = \frac{1}{2\pi} |\textbf{h}|^{-\frac{1}{2}} e^{-\frac{1}{2}\textbf{X}^T \textbf{h}^{-1} \textbf{X}}\label{eq:gaussian}
\end{equation} In two dimensions $\textbf{h}$ is supposed to be a $2\times2$ matrix that would contain a vector of bandwidths for the two dimensions to control the amount and orientation of smoothness. However, following the R adaptation of two dimensional kernel density estimator we use a scalar value as the bandwidth $h$ that was taken to apply to both directions. From the estimated function $\rho_k (\textbf{X})$, we finally estimate the density at each of the pedestrian positions and associate the density with the corresponding pedestrian. In Figure 6 the density $\rho_k$ fields of the pedestrians for a typical case in our data set has been shown for two typical values of bandwidth $h$. \begin{figure}[h!]
    \centering
    \includegraphics[width=\linewidth]{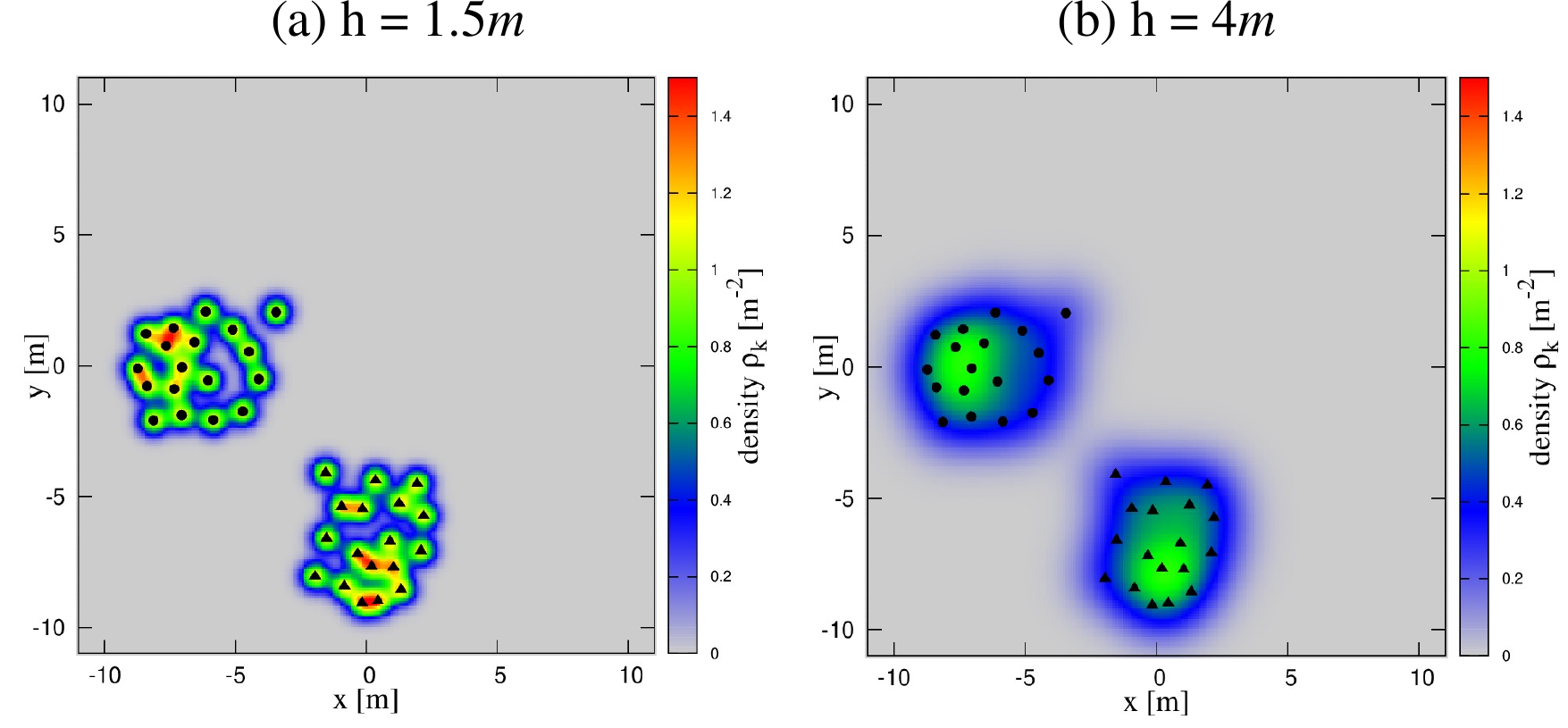}
    \caption{Density $\rho_k$ fields as estimated using the kernel method for two typical values of the bandwidth, viz. (a) $h = 1.5$m and (b) $h = 4$m. Density values decrease with the increase in $h$. The data used here is the same one that has been used in Figures \ref{fig:fields_grid} and \ref{fig:fields_xt}.}
    \label{fig:fields_kernel}
\end{figure}

\subsubsection{Voronoi Method}\label{voronoi}

The Voronoi cell of a pedestrian $P$ is the area $A$ of the surface within which all the points are closer to $P$ than to any other pedestrian $P'$. The density estimate of a pedestrian is then \begin{equation}
    \rho_v=\frac{1}{A}.\label{eq_plain_voronoi}
\end{equation} If we directly use this definition to construct the Voronoi diagram of the pedestrians at every instant of the trial, we notice that pedestrians on the edge of the groups may have a large and possibly infinite Voronoi cell leading to a density near zero. While this indicates that these pedestrians are not constrained by neighbors at least on one side, it may be more appropriate to associate with them the density on the group side.

If the group was confined in a space limited by physical boundaries, the latter could be used to bound the Voronoi cells. In the absence of such boundaries, a possibility is to bound Voronoi cells with an arbitrary limit. For example a restriction to $2$m$^2$ was used in \citep{seyfried2010}, but for a wide variety of real-life situations there is no justified unique choice of this restriction. Here we prefer a method that depends on the density on the group side.

\begin{figure}
    \centering
    \includegraphics[width=\linewidth]{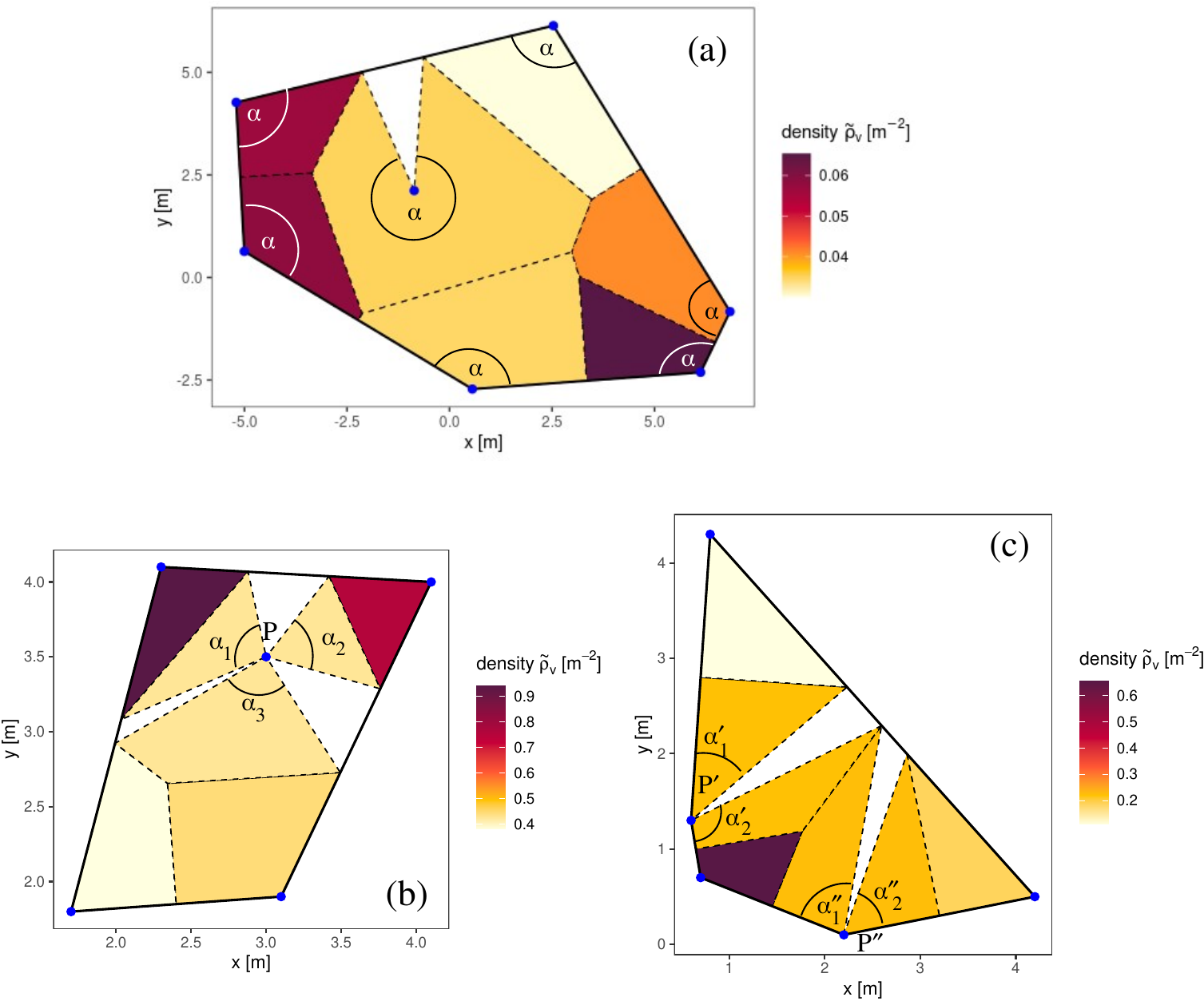}
    \caption{Examples of angular modifications in pedestrian groups. In each of the above panels the data correspond to randomly generated pedestrian positions, denoted by blue dots. The solid line in black denotes the convex hull for the group of pedestrians, and the dashed black lines indicate the Voronoi-cell boundaries. (a) A typical case with 7 pedestrians, where the angular sectors denoted by $\alpha$ are shown for each Voronoi-cell. (b) A typical case with 5 pedestrians, where pedestrian $P$'s Voronoi cell is clipped by the convex hull in 3 directions, leading to multiple angular sectors, where $\alpha=\alpha_1+\alpha_2+\alpha_3$. (c) A typical case with 5 pedestrians, where pedestrians $P'$ and $P''$ lie on the convex hull, and their Voronoi-cells are clipped by the convex hull in the opposite direction. This results in two angular sectors for each case, with $\alpha=\alpha'_1+\alpha'_2$ and $\alpha=\alpha''_1+\alpha''_2$.}
    \label{fig:voronoi3}
\end{figure}

To do this, a first correction proposed in \citep{cecile_vor} is to restrict the Voronoi cells to the angular sectors in which the Voronoi cell lies inside the convex hull encompassing the whole set of pedestrians. In Figure \ref{fig:voronoi3}(a) we have demonstrated a typical case for a randomly generated pedestrian group where the angular sectors, denoted by $\alpha$, for each Voronoi-cell are shown.  A correction in the density accounting for the suppressed angular sectors $2\pi-\alpha$ must be performed as \begin{equation}
    \Tilde{\rho}_v=\frac{\alpha}{2\pi}\frac{1}{A}\label{eq:vor_cecile}
\end{equation} where $\alpha$ is the angular sector on which the Voronoi cell is defined. This takes care of the angular adjustments for the agents located `on’ the convex hull and the agents whose Voronoi cell extend beyond the convex hull. The density values shown in Figure \ref{fig:voronoi3}(a) are estimated by using Eq. (\ref{eq:vor_cecile}).

However, for very small samples as the ones in our data set, there are some specific cases that have to be taken into account. For example, a Voronoi cell can extend on multiple sides of the group, requiring to suppress multiple sectors. One such example for a typical data set from our experiments is illustrated in Figure . \begin{figure}[h!]
    \centering
    \includegraphics[width=\linewidth]{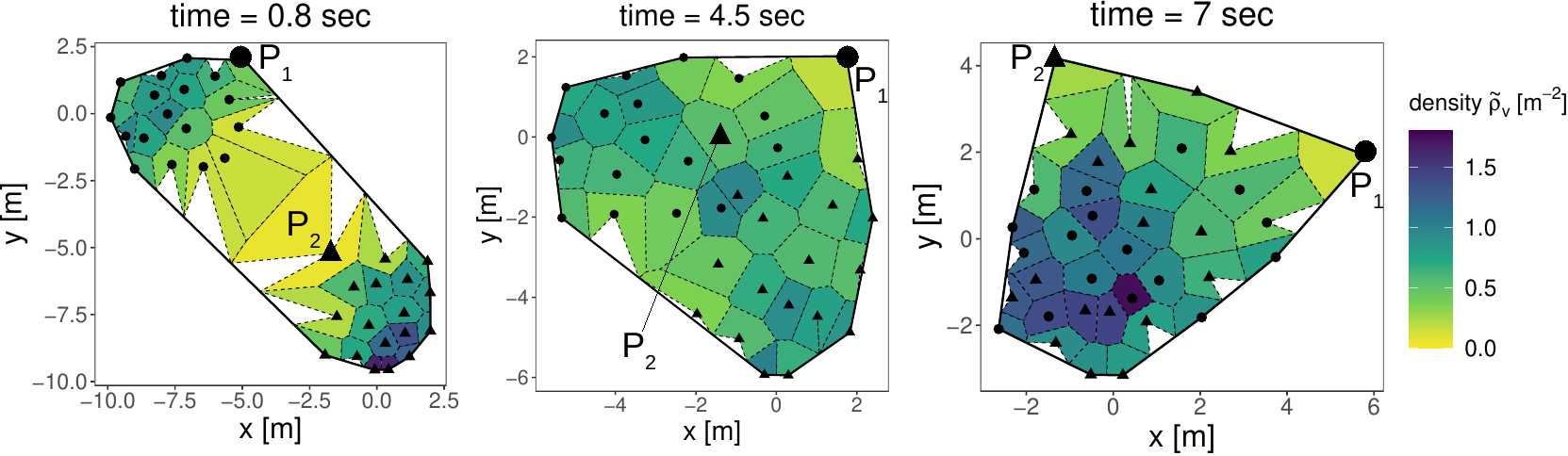}
    \caption{Modified Voronoi cells for a trial with a $90\degree$ crossing angle at three different instances. The black line represents the current convex hull. Dots and triangles in black indicate pedestrians from two groups, moving along the $x$-axis and $y$-axis respectively. Dotted lines mark the Voronoi cell boundaries. The entire video of this experimental trial with modified Voronoi cells is available as a supplementary material (S1 Video). The time-sequence of density estimations for pedestrians denoted by $P_1$ (bigger black dot) and $P_2$ (bigger black triangle) are shown in Figure \ref{fig:rho_vs_rhovc}. Pedestrian $P_2$ at time 0.8 sec is an example whose Voronoi cell extends in 2 directions to the convex hull, requiring suppression of two angular sectors. The entire video of this trial along with the modified Voronoi cells is provided as a supplemental online material (S1 Video).}
    \label{fig:voronoi3_data}
\end{figure} The Voronoi cell corresponding to one of the pedestrians ($P_2$ at $0.8$ sec) subtends 2 angular sectors with the convex hull. For such cases, Equation (\ref{eq:vor_cecile}) can not be applied to find the density.

For even smaller group of pedestrians, the number of such multiple sectors could increase. In Figure \ref{fig:voronoi3}(b) we show a randomly generated pedestrian group of 5 where the Voronoi cell of a pedestrian clipped by the convex hull extends along 3 directions. Another typical case for smaller pedestrian groups could be a pedestrian lying on the convex hull itself, but its Voronoi cell extends beyond the convex hull in the other direction, as shown in Figure \ref{fig:voronoi3}(c). For both the cases shown in Figures \ref{fig:voronoi3}(b) and \ref{fig:voronoi3}(c), Eq. (\ref{eq:vor_cecile}) becomes invalid for density estimation.

For these cases in which multiple angular sectors must be suppressed, we propose the following modification of Equation (\ref{eq:vor_cecile}): as \begin{equation}
    \tilde{\rho}_v = \frac{\sum_i\alpha_i}{2\pi} \frac{1}{A}\label{eq_vor_mod}
\end{equation} where $i(> 1)$ is the number of problematic angular sectors that are clipped, as described above. The density values shown in Figures. \ref{fig:voronoi3}(b) and \ref{fig:voronoi3}(c) were estimated by applying Equation (\ref{eq_vor_mod}). Notably, when there are multiple angular sectors to consider, application of Equation (\ref{eq:vor_cecile}) would not be possible at all to estimate density, and Equation (\ref{eq_vor_mod}) has to be included in the computational algorithm.

Although the data set that we are primarily working with has $36 - 38$ pedestrians, during the time evolution of the two groups crossing each other, or in any other pedestrian situation in general, such peculiar cases occur and our computational strategy pertains to them. The code used for the Voronoi method with angular modifications is also available in the public repository \url{https://doi.org/10.5281/zenodo.8138327}, along with the crossing flows data set \url{https://doi.org/10.5281/zenodo.5718430}.

\section{Results and Discussion}\label{sec:results}

\begin{figure}[h!]
    \centering
    \includegraphics[width=\linewidth]{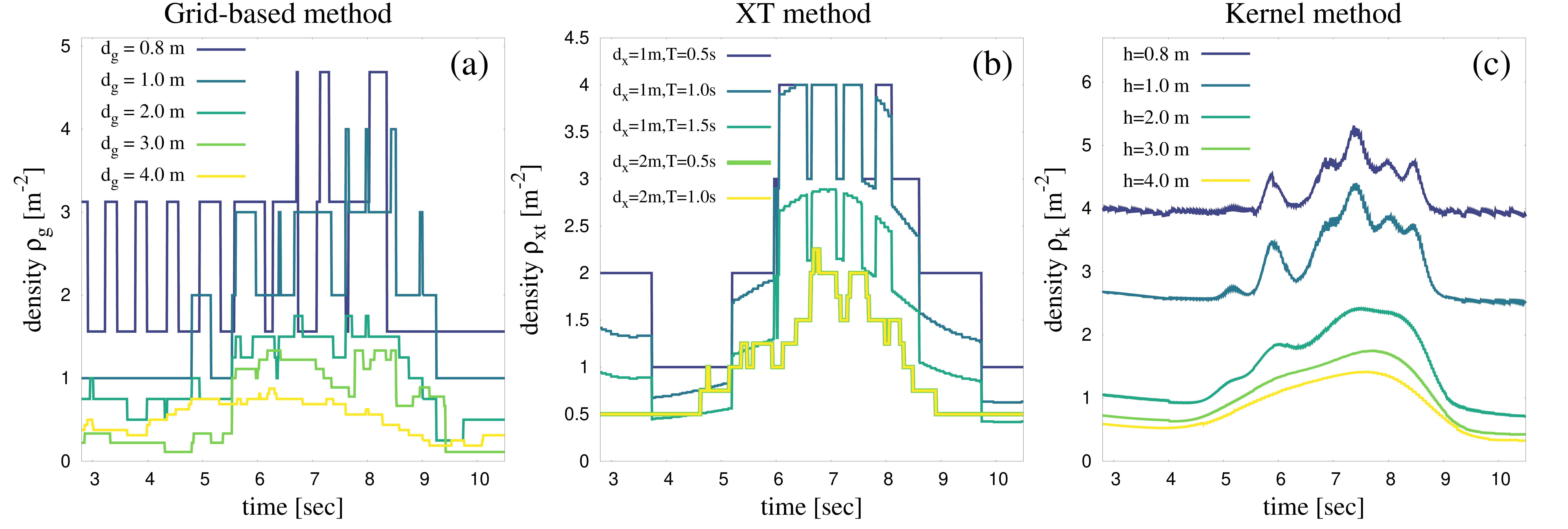}
    \caption{Temporal variations of density as a function of time, along the trajectory of a given pedestrian. Density is determined by (a) the classical grid-based method, (b) the XT-method, (c) the Gaussian kernel method. Time sequences are shown for several values of the spatial scale $d_g$, $d_x$, or $h$ and in (b), of the time window $T$.}
    \label{fig:density_samples}
\end{figure}

In the first three methods described in the previous section, the parameters $d_g$, $d_x$, or $h$ determine the typical spatial length on which the presence of a pedestrian has an effect on the density when measuring the density field. It also determines the density value at the pedestrian location when density is measured along a trajectory. In Figure \ref{fig:density_samples} we show how the density along the trajectory of a given pedestrian measured by the 3 aforementioned methods varies as the spatial parameters $d_g$, $d_x$, or $h$   decrease.  We shall see in this section that, while some methods estimate the pedestrian density without any difficulty, others completely fail once they are used to determine density along a trajectory.

\begin{figure}[h!]
    \centering
    \includegraphics[width=\linewidth]{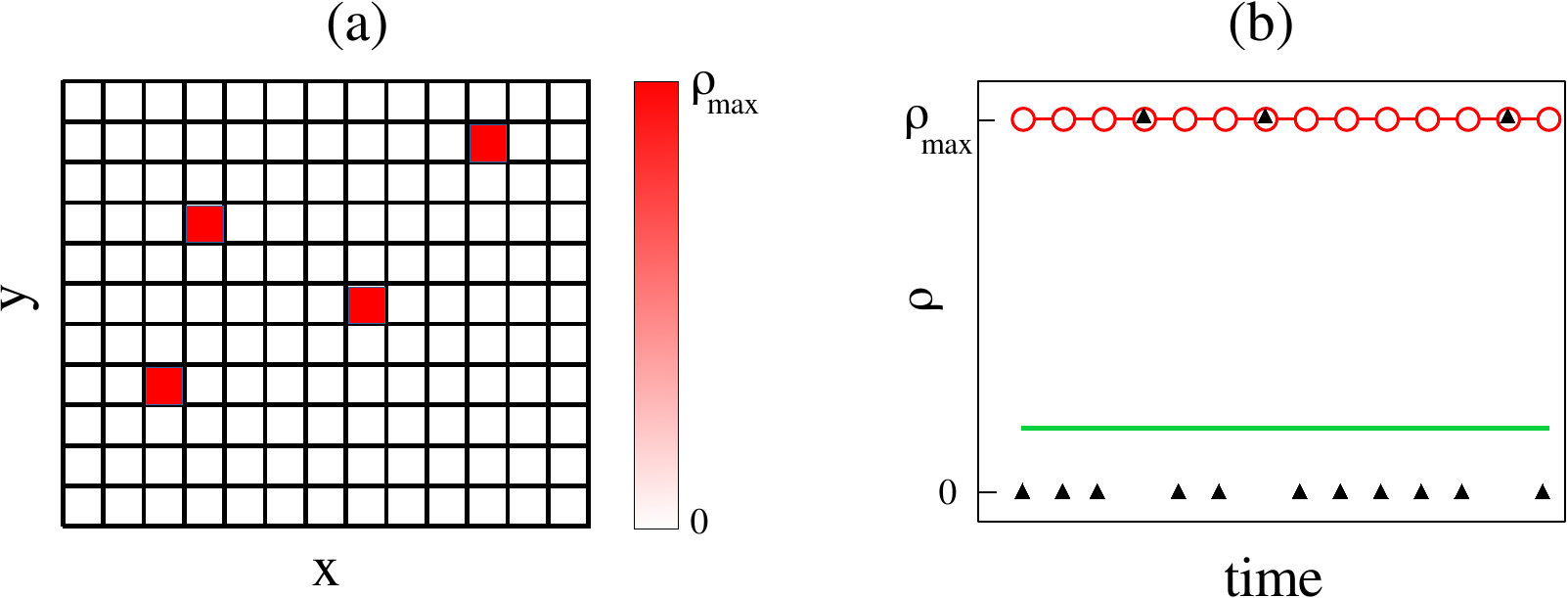}
    \caption{Sketch representing the variation in density estimated by the grid-based classical method (a) when the whole density field is plotted, (b) for a time evolution in a fixed cell (black triangles) or along a pedestrian trajectory (red circles). Solid line in red and green indicate the average values.}
    \label{fig:grid_sketch}
\end{figure}

We first illustrate how this failure occurs using a classical grid-based method. As we are interested in situations with small number of pedestrians, $d_g$ must be small enough to capture the immediate vicinity of a given pedestrian. In the limit of small cells, at most one pedestrian can be contained in the cell. The corresponding density is $\rho_\text{max} \equiv 1/d_g^2$ . If we plot the whole density field, we get typically a figure as in Figure \ref{fig:grid_sketch}(a) with some empty cells and some cells with density $\rho_\text{max}$.  The density field is not smooth, but using a spatial average, we get a reasonable density estimate - provided the area on which we average is homogeneous. For stationary flows, it is also possible to average in time even for a rather small measurement area \citep{zhang2011} and to get average density values with rather low fluctuations.

Now instead, if we decide to measure the density along the trajectory of a given pedestrian, the density in the cell where this pedestrian is located will always be $\rho_\text{max}$ - a value which can be arbitrarily high. Any kind of averaging will always give $\rho_\text{max}$ (red line in Figure \ref{fig:grid_sketch}(b)), which is very far from the real average density value found previously (green line in Figure \ref{fig:grid_sketch}(b)). Actually density values are completely biased by the fact that we have selected cells conditionally, along the pedestrian trajectory.

Another variant could be to measure the density along the trajectory of a pedestrian without counting the pedestrian himself. But then again we encounter a bias: for small cells, the density along the trajectory would always be zero, far from the real density value. This grid-based method could be used to estimate the local density around the pedestrian trajectory only if the cell size could be chosen large compared to the inter-pedestrian distance, but small compared to density gradients. This is obviously not possible for the small pedestrian groups, similar to the ones involved in our experiment.

A similar failure for density measurements along trajectories can be observed for the next two methods as well, viz. XT method and Kernel method. The systematic bias and the fluctuations around this biased value have been studied in detail for one dimensional systems in \citep{tordeux2015}.

So to summarise for the first three methods, the effect of decreasing the spatial parameters $d_g$, $d_x$, or $h$ is two-fold: fluctuations increase, and the average value increases towards arbitrarily high values. While the increase of fluctuations could possibly be reduced by more averaging, the bias introduced in the average value cannot be avoided and makes these methods ill-defined for a measurement along a pedestrian trajectory.

\begin{figure}[h!]
    \centering
    \includegraphics[width=\linewidth]{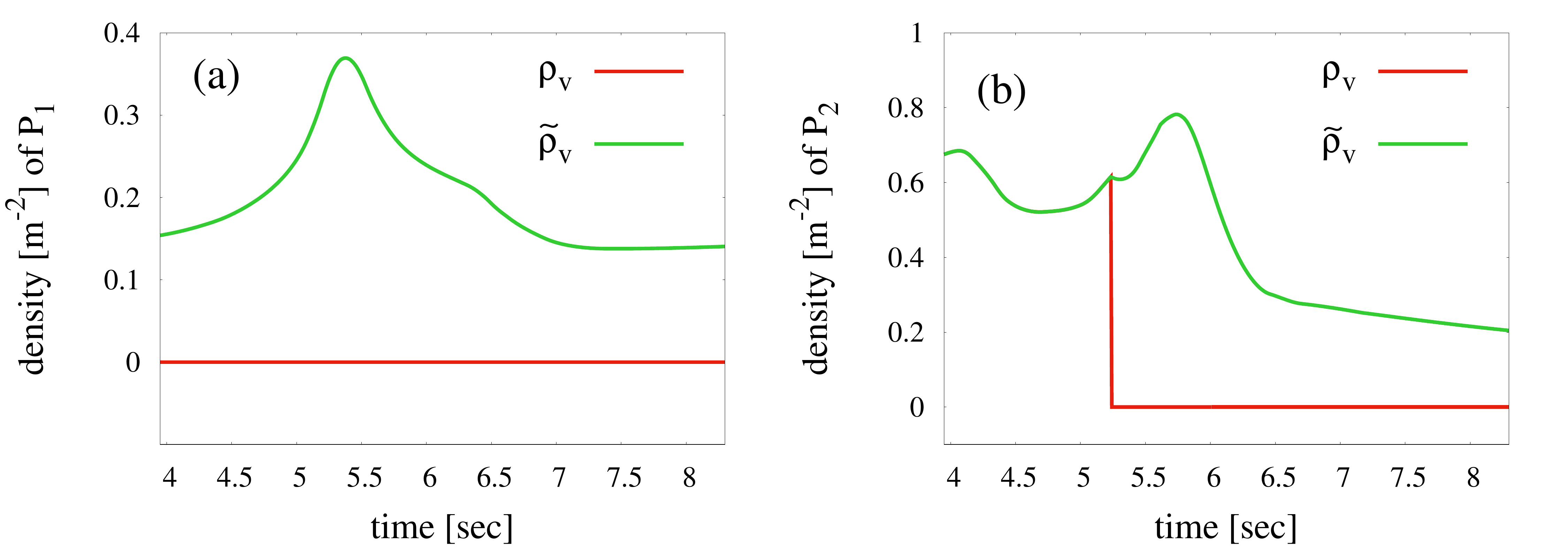}
    \caption{Time sequences of the density $\rho_v$ (considering original Voronoi method) and $\Tilde{\rho}_v$ (considering modified Voronoi cells clipped within the convex hull) for two typical pedestrians, (a) $P_1$ and (b) $P_2$ , denoted by the bigger black circle and triangle respectively in Figure \ref{fig:voronoi3_data}.}
    \label{fig:rho_vs_rhovc}
\end{figure}

On the other hand, there are some methods that do not have the problem of bias along pedestrian trajectories as we discussed so far. These are the methods which, by definition, scan the surroundings of a pedestrian on a scale such that the nearest neighbors’ positions can be taken into account. Among the various methods having this property, we here focus on Voronoi method - a method which is very popular due precisely to its good behavior in a large range of situations \citep{seyfried2010,zhang2011,DUIVES2015162,cao2017,cecile_vor}, in spite of the fact that some fluctuations are introduced by the piecewise constant nature of the density field. In the previous section we have described how this method can be adapted to account for small system sizes. In Figure \ref{fig:rho_vs_rhovc} the variations of $\rho_v$ (density obtained by original Voronoi method, Equation (\ref{eq_plain_voronoi})) and $\Tilde{\rho}_v$ (considering modified Voronoi cells clipped within the convex hull with sector suppression, Equation (\ref{eq_vor_mod})) as a function of time have been plotted for 2 typical pedestrians. 

It is worth mentioning here that density $\rho_v$, as defined by Equation (\ref{eq_plain_voronoi}), is apparently meaningless for agents on the edge of the groups, unless there are physical boundaries or at least imaginary boundary, such as the convex hull. The reason for this is quite obvious, which is that an infinitely large Voronoi cell of these agents would make the density value equal to zero, which however is not physical. So in the literature of pedestrian density estimation, whenever the Voronoi cell based method is employed, some bounding line has been used, even though arbitrary, to limit the infinitely large Voronoi cells near the boundary \citep{helbing2007pre}. Otherwise, one is restricted to consider only the agents who are well within the bulk, i.e., whose Voronoi cells have no effect whether clipped or not.

From Figure \ref{fig:rho_vs_rhovc}(a) we can see that for pedestrian $P_1$ in Figure \ref{fig:voronoi3_data}, the density $\rho_v$ by original Voronoi method remains zero, which which does not account for the fact that $P_1$ is surrounded by pedestrians on one side. Whereas, the density $\tilde{\rho}_v$ obtained by clipping the Voronoi cells by convex hull and including the angular sector corrections, produces a realistic estimate of density in the pedestrian’s neighborhood.

On the other hand, for pedestrian $P_2$ in Figure \ref{fig:voronoi3_data} we get an interesting observation. The density estimate, as could be seen in Figure \ref{fig:rho_vs_rhovc}(b), remains the same for $\rho_v$ and $\tilde{\rho}_v$ until about $5.25$ sec. After this, $\rho_v$ suddenly drops to zero, but $\tilde{\rho}_v$ continues to provide a realistic estimate of density. This can be understood from Figure \ref{fig:voronoi3_data}. We see that for pedestrian $P_2$, the Voronoi cell remains well within the bulk at time $4.5$ sec. As time progresses, this pedestrian moves upwards and remains located on the convex hull for the rest of the trial, naturally subtending an infinite Voronoi cell. It is only by clipping the cell by the convex hull and including the angular corrections that we obtain a physically possible density value. Hence the improvement in density estimation is brought about by $\tilde{\rho}_v$.

\section{Conclusion}\label{sec:conclu}

In this paper we have presented a new computational algorithm based on Voronoi cells that makes it possible to estimate the density for individuals in the case of small groups moving without a spatial boundary. We begin by presenting an evaluation of the existing methods employed for density estimation in similar contexts, which highlights the limitations and drawbacks of these approaches when calculating the density along the trajectory of a chosen pedestrian. The fact that these methods are parameter dependent makes them ill-defined for realistic, stable estimation of density in an individual’s neighborhood. The density obtained using these methods contains systematic bias and fluctuations depending on the chosen parameter value. We then consider a Voronoi cell based method of density estimation \citep{cecile_vor} and propose a modification (Figure \ref{fig:voronoi3}) for case of small pedestrian groups to obtain an unbiased, stable, parameter-independent estimate of the density. The resulting technique can be applied to small or large crowds, with or without physical boundaries.

In the literature on pedestrian density estimation, the XT method is often considered a rigorous method for constructing fundamental diagrams. However, calculating the pedestrian density using the XT method has several limitations. To begin with, an instantaneous measurement of density is not possible in the XT method. We see that the value of density at time t, actually depends not only on the current state of the system, but also on its state in the past and in the future. Specifically, the trajectory of the pedestrian under consideration, and those of its neighbors have to be taken into account within the time interval $T$, not just their positions at time $t$. This is a problem for highly non-stationary dynamics as in the crossing events considered here.

By contrast, in the Voronoi cell based density estimation, only the positions of pedestrians at time $t$ are needed to calculate the instantaneous density of the neighborhood. The availability of data with longer trajectories is not essential for the Voronoi method to be useful.

Another drawback of the XT method for density estimation is its dependence on the parameters $d_x$ and $T$. In our paper, we have demonstrated that density measurements using the XT method fluctuate significantly depending on the chosen parameters. There are no universally accepted values of $d_x$ and $T$ that should be used for density estimations. On the other hand, the advantages that we get by using Voronoi cell based density estimation is that it does not depend on the choice of parameter values, gives us a stable and realistic estimate of density in an individual’s neighborhood, and has physical meaning for a wide variety of human crowd situations.

Our proposed method of pedestrian density estimation could facilitate the construction of fundamental diagrams (FD) from the point of view of individual pedestrians, along their trajectories. Thanks to our method, one is not restricted to consider the collective density of the whole group to study the fundamental relation between density and velocity of the flow. This microscopic approach could potentially lead to the creation of more realistic models of crowd simulation, followed by more efficient methods of crowd management. In our next endeavor we intend to study FDs for various conditions of the present data set. Several other methods can also allow to measure density along a trajectory, for example using the harmonic sum of distances from a pedestrian to neighbors within the pedestrian’s field of view \citep{DUIVES2015162}, or considering the region where the convex hulls of the two crossing groups have an intersection, etc.

More generally, one has to keep in mind that at small scales, a density cannot be defined in a strict way as for example for a fluid. Indeed, a proper definition of density requires that there is a scale separation between, on the one hand, the individual scale and the dimension of the area in which density is computed by a proper averaging, and on the other hand, the scale on which density gradients occur, or on which boundary effects become important. This is rarely the case for pedestrians, and even less for the small pedestrian
groups that we consider here.

What we call density is thus rather an observable that we think is relevant for the navigation of pedestrians, and that we could call perceived density, following \citep{seyfried2010}. It certainly depends on neighboring pedestrians, but could depend more specifically on their relative positions (rear or front), on possible visual occlusions \citep{greg2022}, on interpersonal distances \citep{geoerg2022}, etc. One must thus keep in mind that the choice of the perceived density plotted in the fundamental diagram is actually a hypothesis on the driving cue determining pedestrian dynamics. We shall explore various choices for this observable in our future research.\\

\noindent\textbf{Supplemental online material}\\

\noindent\textbf{S1 Video.} Video of the modified Voronoi cells for all the pedestrians involved in the crossing flows the experimental trial shown in Figure \ref{fig:voronoi3_data}, where we have shown snapshots of this video for 3 typical instances.\\

\noindent\textbf{Funding}\\

PM acknowledges financial support from Bretagne S.A.D., France and National Science Center (NCN, Poland) through SONATA grant no. 2022/47/D/HS4/02576.\\

\noindent\textbf{Disclosure Statement}\\

The authors report that there are no competing interests to declare.\\

\noindent\textbf{Data availability statement}\\

The data set for crossing flows of pedestrians, which has been used in this research is available at \url{https://doi.org/10.5281/zenodo.5718430}. The computational method (in R) to use Voronoi cells with angular modifications to estimate individual densities in small pedestrian groups moving without physical boundaries is available at \url{https://doi.org/10.5281/zenodo.8138327}.\\
\bibliographystyle{elsarticle-num} 
\bibliography{cas-refs}





\end{document}